\documentclass[twoside,twocolumn,9pt]{article}
\usepackage{extsizes}
\usepackage[super,sort&compress,comma]{natbib} 
\usepackage[version=3]{mhchem}
\usepackage[left=1.5cm, right=1.5cm, top=1.785cm, bottom=2.0cm]{geometry}
\usepackage{balance}
\usepackage{times,mathptmx}
\usepackage{sectsty}
\usepackage{graphicx} 
\usepackage{lastpage}
\usepackage[format=plain,justification=justified,singlelinecheck=false,font={stretch=1.125,small,sf},labelfont=bf,labelsep=space]{caption}
\usepackage{float}
\usepackage{fancyhdr}
\usepackage{fnpos}
\usepackage[english]{babel}
\addto{\captionsenglish}{%
  
}
\usepackage{array}
\usepackage{droidsans}
\usepackage{charter}
\usepackage[T1]{fontenc}
\usepackage[usenames,dvipsnames]{xcolor}
\usepackage{setspace}
\usepackage[compact]{titlesec}
\usepackage{hyperref}

\usepackage{epstopdf}

\definecolor{cream}{RGB}{222,217,201}

\begin{document}

\pagestyle{fancy}
\thispagestyle{plain}
\fancypagestyle{plain}{

\fancyhead[C]{\includegraphics[width=18.5cm]{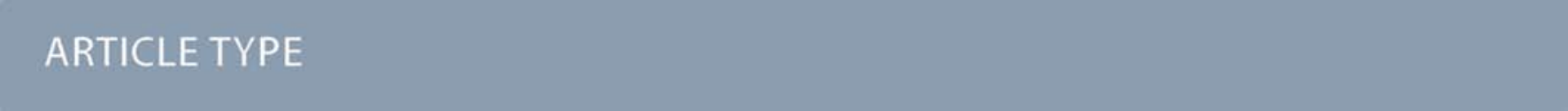}}
\fancyhead[L]{\hspace{0cm}\vspace{1.5cm}\includegraphics[height=30pt]{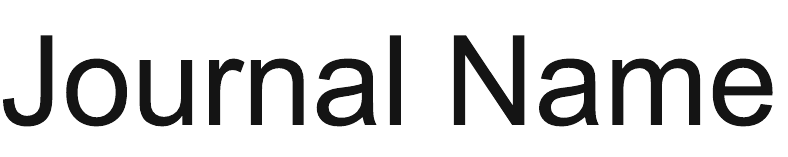}}
\fancyhead[R]{\hspace{0cm}\vspace{1.7cm}\includegraphics[height=55pt]{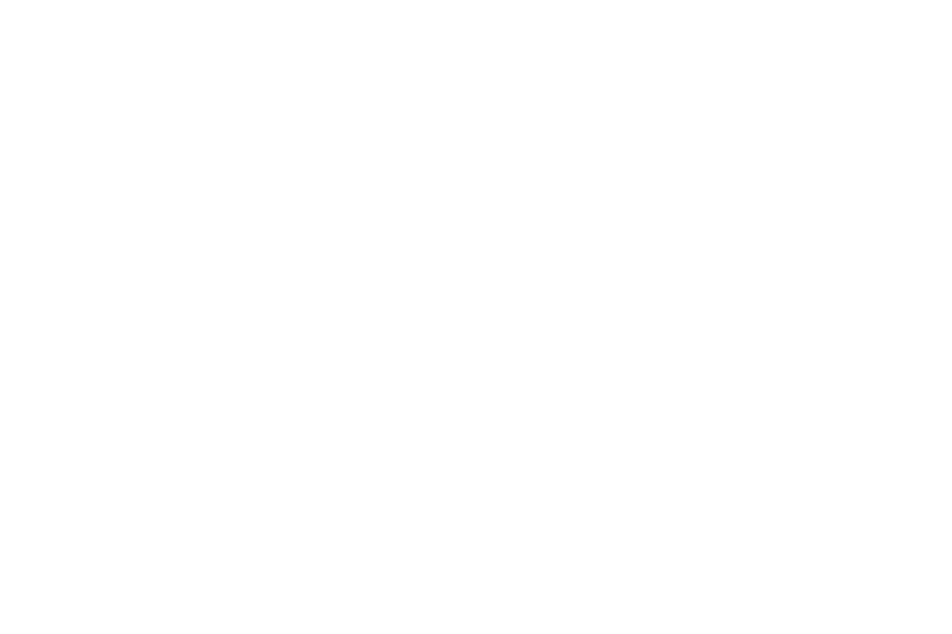}}
\renewcommand{\headrulewidth}{0pt}
}

\makeFNbottom
\makeatletter
\renewcommand\LARGE{\@setfontsize\LARGE{15pt}{17}}
\renewcommand\Large{\@setfontsize\Large{12pt}{14}}
\renewcommand\large{\@setfontsize\large{10pt}{12}}
\renewcommand\footnotesize{\@setfontsize\footnotesize{7pt}{10}}
\makeatother

\renewcommand{\thefootnote}{\fnsymbol{footnote}}
\renewcommand\footnoterule{\vspace*{1pt}%
\color{cream}\hrule width 3.5in height 0.4pt \color{black}\vspace*{5pt}} 
\setcounter{secnumdepth}{5}

\makeatletter 
\renewcommand\@biblabel[1]{#1}            
\renewcommand\@makefntext[1]%
{\noindent\makebox[0pt][r]{\@thefnmark\,}#1}
\makeatother 
\renewcommand{\figurename}{\small{Fig.}~}
\sectionfont{\sffamily\Large}
\subsectionfont{\normalsize}
\subsubsectionfont{\bf}
\setstretch{1.125} 
\setlength{\skip\footins}{0.8cm}
\setlength{\footnotesep}{0.25cm}
\setlength{\jot}{10pt}
\titlespacing*{\section}{0pt}{4pt}{4pt}
\titlespacing*{\subsection}{0pt}{15pt}{1pt}

\fancyfoot{}
\fancyfoot[LO,RE]{\vspace{-7.1pt}\includegraphics[height=9pt]{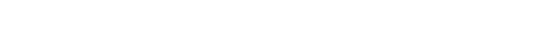}}
\fancyfoot[CO]{\vspace{-7.1pt}\hspace{13.2cm}\includegraphics{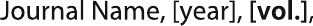}}
\fancyfoot[CE]{\vspace{-7.2pt}\hspace{-14.2cm}\includegraphics{head_foot/RF}}
\fancyfoot[RO]{\footnotesize{\sffamily{1--\pageref{LastPage} ~\textbar  \hspace{2pt}\thepage}}}
\fancyfoot[LE]{\footnotesize{\sffamily{\thepage~\textbar\hspace{3.45cm} 1--\pageref{LastPage}}}}
\fancyhead{}
\renewcommand{\headrulewidth}{0pt} 
\renewcommand{\footrulewidth}{0pt}
\setlength{\arrayrulewidth}{1pt}
\setlength{\columnsep}{6.5mm}
\setlength\bibsep{1pt}

\makeatletter 
\newlength{\figrulesep} 
\setlength{\figrulesep}{0.5\textfloatsep} 

\newcommand{\topfigrule}{\vspace*{-1pt}%
\noindent{\color{cream}\rule[-\figrulesep]{\columnwidth}{1.5pt}} }

\newcommand{\botfigrule}{\vspace*{-2pt}%
\noindent{\color{cream}\rule[\figrulesep]{\columnwidth}{1.5pt}} }

\newcommand{\dblfigrule}{\vspace*{-1pt}%
\noindent{\color{cream}\rule[-\figrulesep]{\textwidth}{1.5pt}} }

\makeatother

\twocolumn[
  \begin{@twocolumnfalse}
\vspace{3cm}
\sffamily
\begin{tabular}{m{4.5cm} p{13.5cm} }

\includegraphics{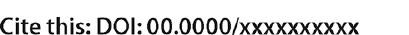} & \noindent\LARGE{\textbf{Surface nanodroplet-based nanoextraction from sub-milliliter volumes of dense suspensions$^\dag$}} \\
\vspace{0.3cm} & \vspace{0.3cm} \\

 & \noindent\large{Jae Bem You,\textit{$^{a,b}$} Detlef Lohse,\textit{$^{b}$} and Xuehua Zhang$^{\ast}$\textit{$^{a,b}$}} \\

\includegraphics{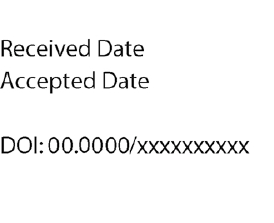} & \noindent\normalsize{Cleaner analytic technique for quantifying compounds in dense suspension is needed for wastewater and environment analysis, chemical or bio-conversion process monitoring, biomedical diagnostics, food quality control among others. In this work, we introduce a green, fast, one-step method called nanoextraction for extraction and detection of target analytes from sub-milliliter dense suspensions using surface nanodroplets without toxic solvents and pre-removal of the solid contents. With nanoextraction, we achieve a limit of detection (LOD) of 10$^{-9} M$ for a fluorescent model analyte obtained from a particle suspension sample. The LOD lower than that in water without particles (10$^{-8} M$), potentially due to the interaction of particles and the analyte. The high particle concentration in the suspension sample thus does not reduce the extraction efficiency, although the extraction process was slowed down up to 5 $min$. As proof of principle, we demonstrate the nanoextraction for quantification of model compounds in wastewater slurry containing 30 wt\% sands and oily components (i.e. heavy oils). The nanoextraction and detection technology developed in this work may be used as fast analytic technologies for complex slurry samples in environment industrial waste, or in biomedical diagnostics.} \\

\end{tabular}

 \end{@twocolumnfalse} \vspace{0.6cm}

  ]

\renewcommand*\rmdefault{bch}\normalfont\upshape
\rmfamily
\section*{}
\vspace{-1cm}


\footnotetext{\textit{$^{a}$~Department of Chemical and Materials Engineering, University of Alberta, Alberta T6G 1H9, Canada; E-mail: xuehua.zhang@ualberta.ca}}
\footnotetext{\textit{$^{b}$~Physics of Fluids Group, Max Planck Center Twente for Complex Fluid Dynamics, JM Burgers Center for Fluid Dynamics, Mesa+, Department of Science and
		Technology, University of Twente, Enschede 7522 NB, The Netherlands.}}

\footnotetext{\dag~Electronic Supplementary Information (ESI) available: [Schematic diagram showing the steps for operating the portable nanoextraction device; plots showing the influence of exposure time on background Nile red fluorescence intensity; limit of detection for Rhodamine B dye]. See DOI: 00.0000/00000000.}



\section{Introduction}
Sample pretreatment is one of the most polluting steps in chemical analysis.\cite{marek2009} In particular, concentrating analytes with an extremely low concentration often requires a large amount of toxic solvents and multiple steps of operation with high energy input for separation such as high speed centrifugation. Waste chemicals is costly to be disposed of safely and sometime poses serious issues to our ecosystem. It is estimated that about 34 million liters of solvent waste are generated annually from liquid chromatography alone, without taking into account the solvent waste produced from sample pretreatment.\cite{marek2016} With the goal of developing sustainable technologies in mind, the solvent waste during sample pretreatment should be minimized.

Development of clean analytic techniques for quantifying traces of compounds from suspensions with high solid concentration of particles is of broad interest in determining environmental pollutants in soil or water,\cite{llompart2019} in process monitoring of fermentation or biomass conversion,\cite{tosi2003,quintelas2019} in biomedical diagnostics,\cite{brassard2019,tabani2019,mashayekhi2010} and in food quality control.\cite{choi2019,zhang2014foodchem} For example, quantification of harmful pesticides (e.g. organophosphates) or pharmaceutical drugs from drinking water treatment sludge is crucial for determining proper disposal procedures.\cite{soares2017} Detection of harmful toxins like cereulide (a type of depsipeptide) or Aflatoxin B$_{1}$ (one of the most dangerous mycotoxins) from food is compulsory to safeguard the public health against foodborne illnesses.\cite{ducrest2019,ren2014}

Presently, removal of the solid contents in such suspension samples prior to analysis of target compounds is required to guarantee high detection sensitivity and to prevent deterioration of analytical instrument, in particular clogging the capillary column in a chromatography system.\cite{boonjob2013} Therefore, development of green analytical technology is needed to simplify the analysis of slurry samples. Specifically, a technology that produce minimal waste, and use non-harmful solvent, as outlined in the 12 principles of green analytical chemistry (GAC),\cite{galuszka2013} is highly desirable.

Currently, solids in suspension samples are mostly removed by centrifugation or filtration. Although these standard methods are simple, some intrinsic limitations hinder efficient and fast extraction from suspensions with high solid contents. First of all, removal of small particles (i.e. $<$ 1 $\mu m$) requires high  power consumption by using specialized centrifugation equipment.\cite{braun2011,hildebrandt2020} Second, clogging of filter pores by the particles is a common issue during filtration,\cite{wang2018clogging, springer2013} potentially producing unnecessary waste. Third, certain target compounds may adsorb on the filter surface reducing extraction efficiency. For instance, pesticides - many of which are hydrophobic - are prone to adsorb on the surface of filter membranes.\cite{kiso2000,doulia2016} Undesired adsorption can lead to the loss of analytes during filtration before detection.\cite{carlson2000,lath2019,ahmad2001} For instance, Ahmad \textit{et al.} reported up to 95\% loss of pesticide analyte from water after syringe filtration.\cite{ahmad2001} Finally, both filtration and centrifugation in the sample pretreatment are not applicable to in-situ analysis of target compounds. An approach to separate, extract and concentrate the analytes directly from slurry samples without removal of solids is needed to significantly simplify and speed up analysis of slurries containing high concentration of solids. 

Enriching traces of target compounds by extraction is an effective way of improving detection sensitivity.\cite{yazdi2010,tang2016} Various forms of extraction have been demonstrated including use of porous membranes in extraction driven by an electric field,\cite{hansen2020,yeh2019,feres2020,stig1999} formation of tiny extractant droplets generated by ultrasonication for liquid phase extraction,\cite{kaw2018,bi2011} or use of nanoparticles as sorbents in solid phase extraction.\cite{bendicho2015,feng2010} In particular, liquid-liquid extraction by using extractant microdroplets can reach high extraction efficiency, due to efficient mass transfer enabled by the high surface area-to-volume ratio of small droplets.\cite{jeannot1996} Approaches such as the single-drop microextraction\cite{jeannot1996,tang2018} or bubble-in-drop microextraction\cite{williams2011,guo2016} use a single extractant droplet hanging at the tip of a syringe needle immersed in the sample solution to extract the target compounds.

The simple setup also enables in-situ analysis in tandem with analytical instruments such as a UV-vis spectrometer.\cite{zaruba2016,zaruba2017} However, the single drop extraction is not compatible with slurry samples as the droplet can easily detach from the needle upon collision with the particles or from the high shear stress generated during the stirring motion.\cite{nancy2006} Although the droplet can be protected with a hollow fibers membrane,\cite{nancy2006,basheer2004} blockage and fouling of pores by the particles can occur and lower the extraction performance.\cite{reyes-garces2018}

Recently, many studies have applied dispersive liquid-liquid microextraction (DLLME) to extract and detect chemicals from various types of samples, including body fluids,\cite{tabani2019,mashayekhi2010} food\cite{boonchiangma2012,zhang2014foodchem,vinas2018}, or environmental samples.\cite{habibollahi2018,ozzeybek2017} In a typical DLLME process, extractant microdroplets form spontaneously by mixing an extractant solution with an aqueous sample.\cite{rezaee2006} Similar to single drop microextraction, target compounds in the mixture are readily extracted into the extractant microdroplets according to high partition coefficient in the droplets. The extractant droplets need to be separated from the mixture and collected by centrifuge. Despite its excellent extraction efficiency and simplicity, DLLME is not suitable for in-situ extraction and analysis of target compounds in solid suspensions as additional filtering or centrifugation steps  are required to remove the solid contents from the sample prior to the extraction procedure. Moreover, environmentally harmful extractants such as halogenated hydrocarbons (e.g. chloroform) are often used.\cite{luo2019,ojeda2018} 

To overcome these drawbacks of DLLME, surface nanodroplets provide an effective approach for liquid-liquid extraction. Surface nanodroplets are droplets on solid surfaces with less than 100 nm in height and sub-femtoliter in volume.\cite{lohse2015revmodphy} Extensive research has shown that surface nanodroplets can be simply produced by solvent exchange process in a controlled way.\cite{zhang2015pnas, wei2020, meng2020, yu2017,lohse2020} Li et al. formed the surface nanodroplets to extract target analytes by flowing the sample solution over the droplets in the same chamber.\cite{li2019small, li2019analchem} In-situ analysis was possible as the nanodroplets are pinned on a substrate during the extraction process. The droplet component was controlled by the solutions used for solvent exchange. Extraction by surface nanodroplets -- i.e. nanoextraction -- also avoids excessive use of chemicals by minimizing the volume of extractant oils and solvents during sample pre-treatment, enabling an eco-friendly analytic technology.

Up to now, nanoextraction has not been applied in analyte concentrating from slurries. It remains unclear how the solids affect the droplet stability and extraction efficiency and rate. In this work, we employ surface nanodroplets to extract and detect compounds from high-solid suspensions --with solid content up to 30 wt\%-- without pre-removal of the solids. Through in-situ detection of a model compound from suspension samples, we find that the final extraction outcome by surface nanodroplets is not influenced by the solid concentration, although the particles may slow down initially due to their adsorption at the droplet interface. As proof-of-concept, we demonstrate extraction and detection of target analyte from industrial waste slurry. The approach reported in our work facilitates and speeds up the pre-treatment of suspension samples by enabling a one-step extraction of target analytes. This novel method may be applicable to extraction and analysis of high-solid suspensions in various fields such as environment monitoring or food quality control. 

\section{Experimental}
\subsection{Hydrophobization of glass capillary for nanodroplet formation}
The surface of glass capillary tubes (Kimble Products, Inc., USA) with an inner diameter of 1.1 $\sim$ 1.4 $mm$, outer diamter of 1.5 $\sim$ 1.8 $mm$, and length of 100 $mm$ was rendered hydrophobic using octadecyltrichlorosilane (OTS) (Sigma Aldrich) by following the protocol reported in ref [\cite{zhang2019ots}]. Briefly, the glass capillary was cleaned with piranha solution made of 70 \% H$_2$SO$_4$ (Fisher Scientific) and 30 \% H$_2$O$_2$ (Fisher Scientific) (v/v) for 15 $min$ at 75 $^{\circ}C$. After cleaning, the capillary tubes were sonicated in water and then in ethanol for 5 $min$ each before drying by a stream of air. Subsequently, the capillary tubes were immersed into an amber bottle containing 100 $mL$ of hexane (Sigma Aldrich) and 100 $\mu$L of OTS. The bottle was tightly closed and kept at room temperature (20.5 $^{\circ}C$) for 12 $hr$. The OTS-treated glass capillaries were then cut to 50 $mm$ in length. Then, the OTS-coated glass capillaries where sonicated in ethanol and in water for 10 $min$ each to remove un-reacted OTS from the surface. 

\begin{figure*}[h]
	\centering
	\includegraphics[scale=1]{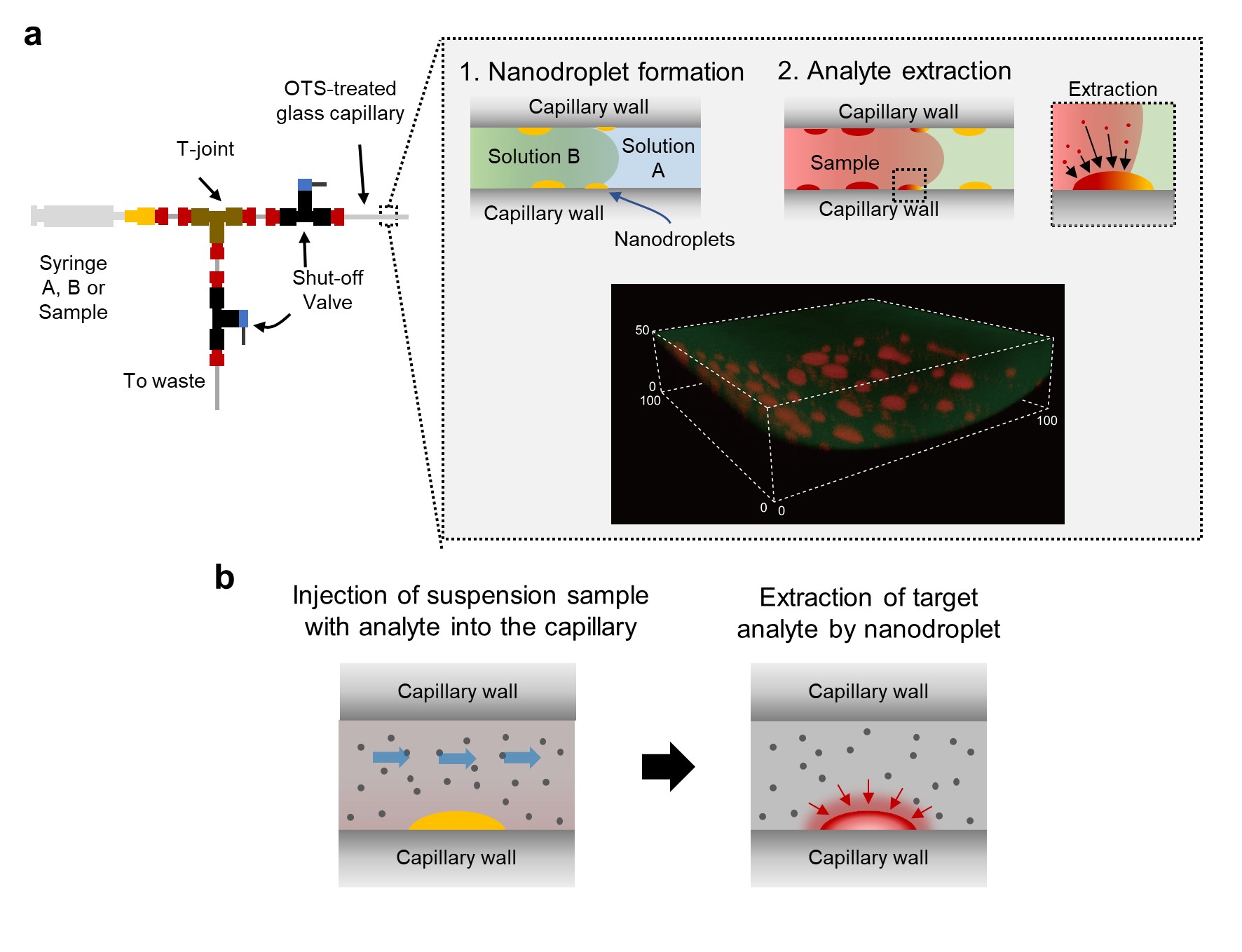}
	\caption{Schematic diagram showing the surface nanodroplet formation and analyte detection using a portable solvent exchange device. a) The surface nanodroplets are formed on the inner wall of OTS-treated glass capillary tube. For detection, the analyte-containing sample can be injected into the capillary tube decorated with surface nanodroplets, into which analytes become extracted. Inset: confocal microsgraph of surface nanodroplets formed on the inner wall of the glass capillary tube. b) Schematic showing extraction of analytes from suspension samples. Extraction is simply achieved by flowing the sample through the capillary tube. As the slurry flows through the capillary tube, analytes are readily extracted into the surface nanodroplets.}
	\label{schematic}
\end{figure*}

\subsection{Formation of surface nanodroplets}
Surface nanodroplets were formed inside the OTS-treated glass capillary tube using the solvent exchange process.\cite{zhang2015pnas} To deliver the solutions, the capillary tube was connected to a portable device composed of two shut-off valves joined by a T-junction (Fig. \ref{schematic}a).\\
\indent The shut-off valves could prevent trapping of air during the sequential delivery of solutions A and B, guaranteeing mixing between the two to drive oversaturation of extractant and form the nanodroplets on the capillary tube wall.\cite{zhang2015pnas}\\
\indent The droplets were generated according to the overall procedure  shown in Fig. S1 of the Electronic Supplementary Information (\dag). First, a solution of 5 \% octanol (Sigma Aldrich) in 50 vol\% ethanol (Sigma Aldrich, reagent grade) aqueous solution (solution A) was injected into the glass capillary. At this point, only the outlet valve was open to let the solution flow to the capillary. Then the syringe was removed from the device and pure water (Solution B) was injected into the device using a new syringe. The air trapped during exchange of syringes could be removed by closing the outlet valve and opening the waste valve to block the passage of air to the capillary and flush it out to the side tube. After removing the air, the waste valve was again closed and outlet valve was opened to let water into the capillary to displace solution A and generate the droplets, as shown by the three-dimensional confocal fluorescent image in Fig. \ref{schematic}a 

\subsection{Nanoxtraction of the analyte using surface nanodroplets}
Target analytes were extracted from the solid suspension simply by injecting the suspension sample into the capillary tube with the nanodroplets on the wall. The solid particles in the suspension sample did not damage the nanodroplets during the extraction process due to pinning effect from the capillary tube wall. After extraction, the analytes in the droplet could be observed in-situ using fluorescence microscopy (Fig. \ref{schematic}b). When the analyte concentration is too high and interferes with the detection, it is possible to remove the excess suspension sample in the capillary tube simply by washing it away with water. However, this was not always necessary within the range of concentrations tested in this work.

\begin{figure*}[h]
	\centering
	\includegraphics[scale=1]{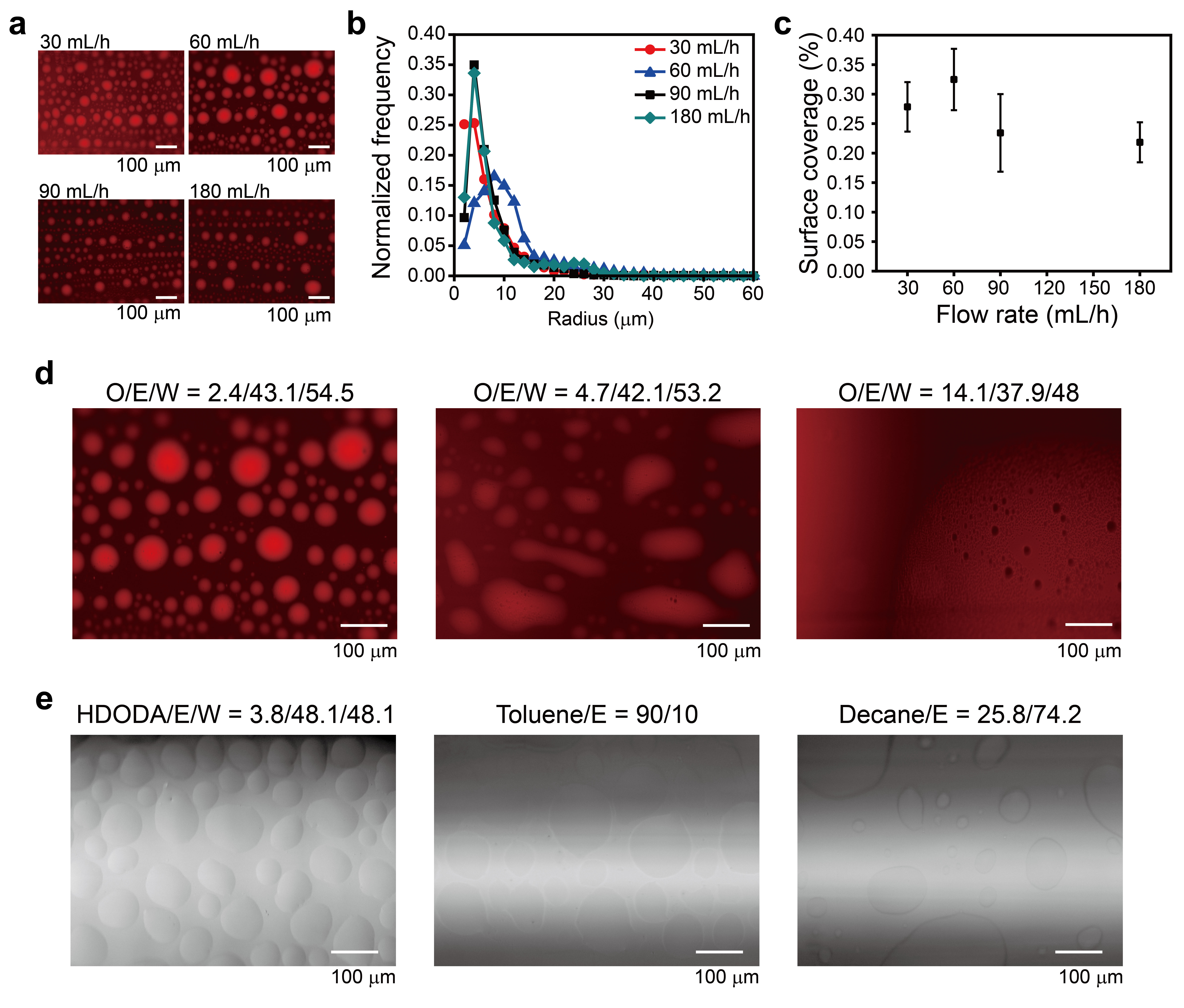}
	\captionsetup{font = {small}}
	\caption{Effect of flow rate, oil concentration and oil type on the formation of surface nanodroplets in the glass capillary tube. a) Fluorescent images of droplets formed with flow rates of 30, 60, 90 and 180 $mL/h$ for solution B. b) Size distribution of droplets formed at each flow rate condition. c) Surface coverage of droplets formed at each flow rate condition (mean $\pm$ S.D.) d) Fluorescent images of droplets formed at different solution A compositions (in wt\%). e) Optical microscope images of droplets formed using different types of oils. The compositions are in wt\%}
	\label{flowrate}
\end{figure*}

\subsection{Fluorescent detection of extracted analytes}
For all fluorescence imaging, the capillary containing the droplets with extracted analytes were observed under green laser light to excite both Rhodamine B and Nile red. Otherwise noted, the exposure times were kept constant. The intensity values of droplets were analyzed using ImageJ. 

For the limit of detection tests, as the range of concentrations of analyte in the suspension sample is large (covering three orders of magnitude), it was not possible to fix an exposure time to measure the fluorescence signals of droplets for all the cases. An exposure time set to detect the analyte concentration corresponding to $10^{-6} M$ was too short to analyze droplets with lower concentrations such as $10^{-9} M$. On the other hand, selecting an exposure time appropriate to detect an analyte concentration of $10^{-9} M$ was too long for a concentration of $10^{-6} M$ and resulted in intensity saturation. Therefore, in order to determine the limit of detection, the images were acquired at different exposure times, and the intensity values were normalized to the value corresponding to 400 $ms$ of exposure time. To avoid erroneous normalization, fluorescent intensities of the background were obtained at various exposure times for different concentrations within the linear region (Fig. S2 in ESI\dag). 

\section{Results and discussion}

\subsection{Formation of surface nanodroplets on the capillary wall for nanoextraction from solid suspensions}

First, we demonstrate how the injection flow rate and the oil concentration influence the formation of surface nanodroplets on the capillary tube wall. The optical images in Fig. \ref{flowrate}a show that the droplets were similar in size and surface coverage on the wall, independent of the flow rate, which ranged from 30 to 180 $mL/h$. At the fastest flow rate, it took only $\sim$ 20 $s$ to form the surface nanodroplets on a 50 $mm$-long tube due to the small diameter and volume of the tube. The probability distribution function (PDF) plots in Fig. \ref{flowrate}b show that most droplets were $\sim$ 5 $\mu m$ in base radius. The surface coverage (Fig. \ref{flowrate}c) was between 20 \% and 35 \%, comparable to the value on a flat substrate with the same coating.\cite{xu2017} We attribute the independence of the droplet size distribution on the flow rate to the geometry confinement in a capillary tube. This is in contrast to the faster droplet growth at a faster flow demonstrated in a flow chamber with a flat substrate.\cite{zhang2015pnas}

An effective way to vary the droplet size is simply changing the oil (i.e. octanol) concentration in solution A. Fig. \ref{flowrate}d shows the fluorescent images of droplets formed with octanol as extractant. The flow rate was kept at 60 $mL/h$ for all cases. As the concentration of octanol in solution A varied from 2.4 wt\% to 4.7 wt\%, the average base radius of the droplets increased from 10.9 $\mu m$ to 14.3 $\mu m$. At even higher concentration of 14.1 wt\%, the droplets were too large, covering the field of view. Controlling the size is important for achieving high extraction efficiency as the droplet volume is known to influence the extraction performance.\cite{jeannot1997}  

Surface nanodroplets can also be formed using many other types of extractants such as 1,6-Hexanediol diacrylate (a type of monomer), toluene (aromatic solvent), and decane (alkane), as shown in the optical microscope images in Fig. \ref{flowrate}e. The selection of solution A and B in each case was guided by the solubility diagram of the extractant liquid, the good solvent, and the poor solvent. In literature, solvent exchange has been applied for forming droplets of unconventional liquids such as extremely viscous silicone elastomers\cite{dyett2017} or ionic liquids.\cite{yu2020} We expect that these droplets can be formed on capillary tubes as well since the principle of droplet nucleation and growth is the same regardless of chamber geometry. Formation of droplets with various extractant liquids expands the type of chemicals that can be extracted and analyzed, as extraction of certain analytes can depend on the extractant type.\cite{wang2014} Moreover, a wide library of extractant liquids avoids the use of toxic and harmful extractants.

Next we demonstrate the capability of surface nanodroplets to extract analytes from solid suspensions. Nile red fluorescent dye at initial concentration of 10$^{-6} M$ was used as a model compound. The dye was dissolved in suspensions containing 3.125 $mg/mL$ to 12.5 $mg/mL$ of 150 $nm$ silica particles. This range of nanoparticles concentration is similar to the solid contents in sewage sludge (10 $mg/kg$ $\sim$ 1000 $mg/kg$).\cite{kim2012} 

\begin{figure*}[h]
	\centering
	\includegraphics[scale = 1]{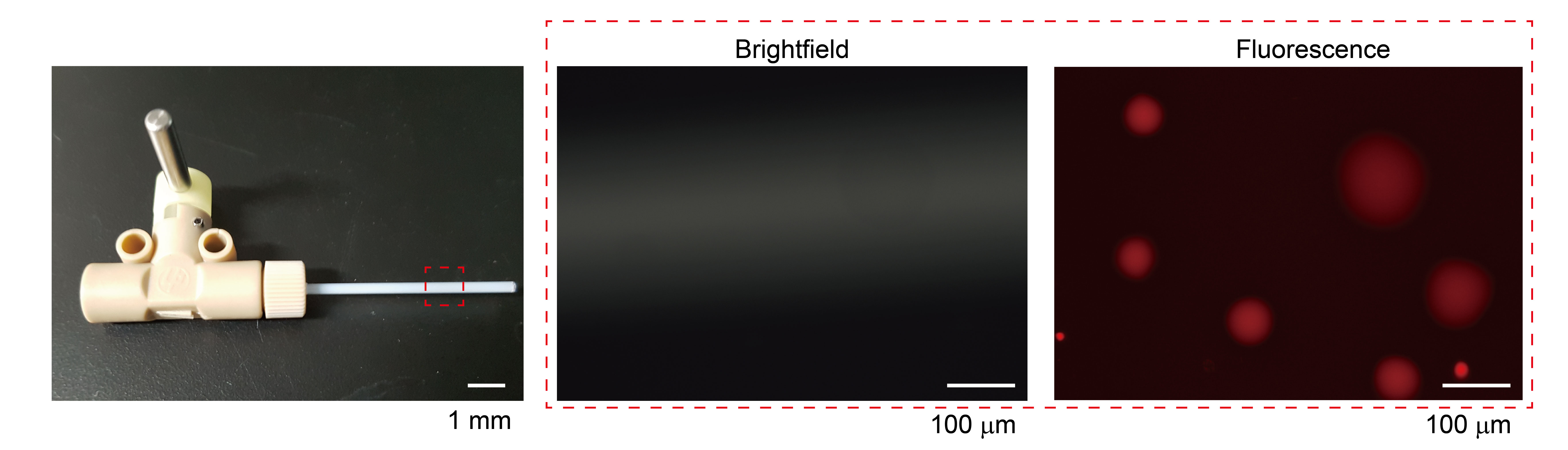}
	\caption{Extraction of Nile red dye from samples with various particle concentrations. Digital camera image of glass capillary filled with silica solution (left). Bright field (center) and fluorescent images of the capillary tube (right). The droplets are not visible in bright field but are clearly shown in the fluorescent image due to extraction of Nile red dye.}
	\label{solid}
\end{figure*}

When the particle suspension was loaded into the capillary tube coated with preformed surface nanodroplets, it was difficult to observe the droplets in bright-field microscope images due to scattering of light by the particles (Fig. \ref{solid}). However, in fluorescent images, the droplets were clearly visible as the model analyte was readily extracted from the suspension into the droplets. The strong fluorescence intensity in the droplets indicates successful extraction of the dye (the target analyte) by the surface nanodroplets even in the presence of solid particles. In the following sections, we show the influence of the solid contents on the nanoextraction kinetics and detection sensitivity.

\subsection{Effect of solid contents on nanoextraction kinetics}
\begin{figure*}[h]
	\centering
	\includegraphics{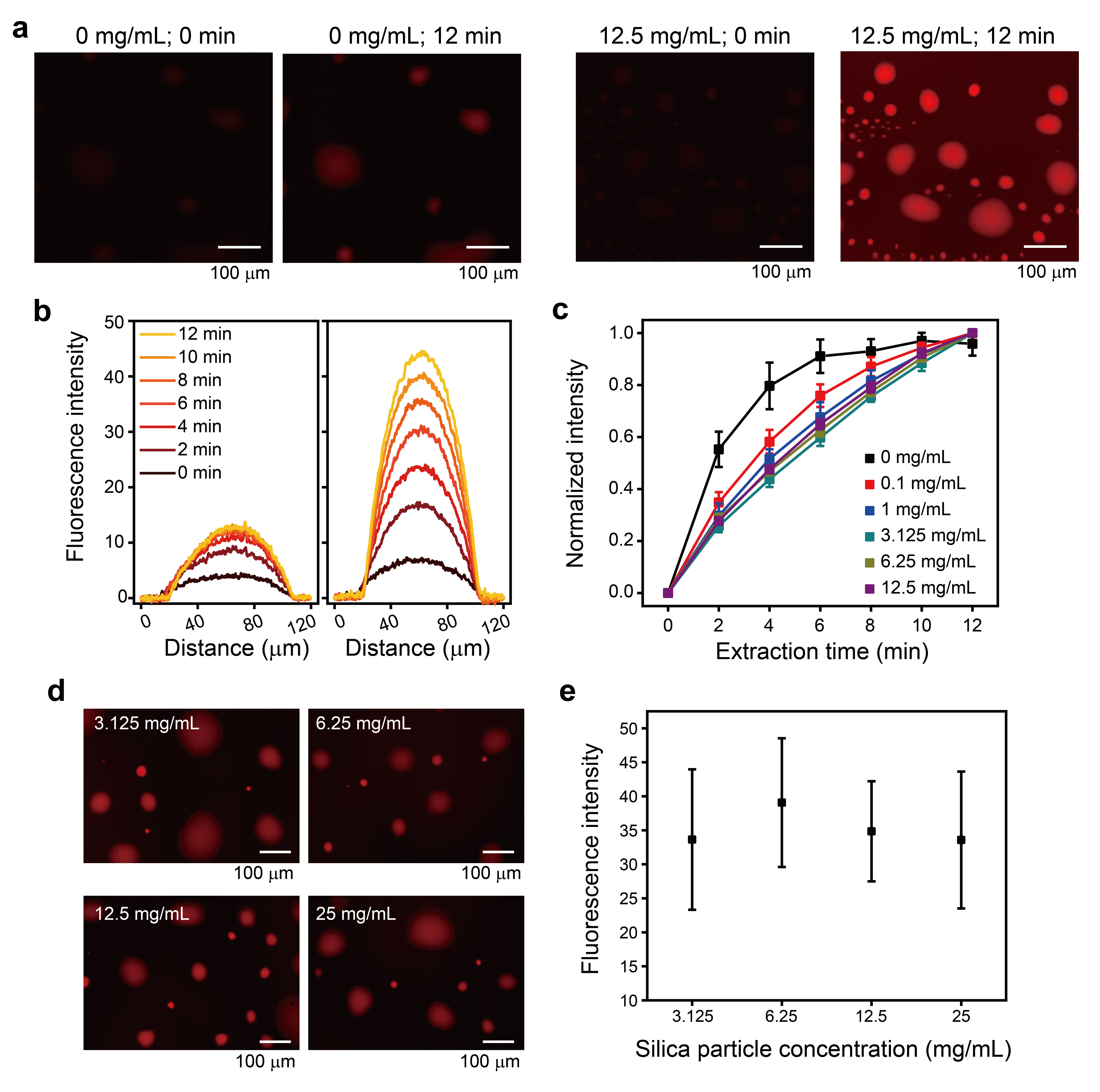}
	\caption{Influence of solid particle on extraction time. a) Fluorescent images of droplets extracting Nile red from sample with no particle (0 $mg/mL$) and with particle (12.5 $mg/mL$). b) Intensity profile across a droplet in samples with and without particle. c) Comparison of normalized maximum intensities of measured droplet over time. 
		d) Fluorescence images and e) intensities of droplets after extracting Nile red from the slurry solutions with particle concentrations of 3.125, 6.25, 12.5, and 25 $mg/mL$, respectively. Here, the Nile red concentration was 10$^{-6} M$ for all cases. Error bars represent $\pm$S.D.}
	\label{extraction_time}
\end{figure*}

Next we compare the nanoextraction kinetics in water and in a suspension sample. Fig. \ref{extraction_time}a compares the fluorescent images of droplets extracting dye from an aqueous solution and from a silica suspension at concentration of 12.5 $mg/mL$. The images at 0 $min$ were acquired as soon as the sample was injected into the capillary tube. Fig. \ref{extraction_time}b shows the intensity profile across a single droplet tracked over 12 $min$. It is clear that for the aqueous solution (i.e. 0 $mg/mL$), the profiles overlap from 6 $min$ whereas that for the silica solution does not reach its maximum value even after 12 $min$.

When the normalized maximum intensity across the droplet was plotted against time (Fig. \ref{extraction_time}c), it was evident that the extraction time is delayed in the case of the suspension sample. The rate was distinguishable up to particle concentration of 0.1 $mg/mL$, but no difference was observed at higher concentrations from 1 $mg/mL$. The slower extraction time in the case of the suspension sample may be attributed to hindered transport of analytes into the droplet by adsorbed silica particles at the droplet-water interface. Similar observations were reported by Goodarzi et al. from liquid-liquid extraction using a single oil droplet rising up a column of particle solutions.\cite{goodarzi2016} Although it has been reported that increased viscosity of a slurry can also reduce mass transfer during extraction,\cite{park2008} this is unlikely the case since the silica concentration used here is not high enough to significantly influence the viscosity.

We can estimate the influence of solids on the extraction kinetics based on the theory established for single drop microextraction.\cite{jeannot1996} The concentration of the analyte extracted into the droplet ($C_{drop}$) is a function of time ($t$), concentration of analyte in the sample ($C_{aq}$), partition coefficient ($p$), the ratio of free droplet interfacial area to droplet volume ($A_{drop}/V_{drop}$), and the mass transfer coefficient ($\bar{\beta}$) relating the diffusivity and concentration boundary layer thickness around the droplet,\cite{jeannot1997}

$$\frac{dC_{drop}}{dt} = \frac{A_{int}}{V_{drop}}\bar{\beta}(pC_{aq}-C_{drop}).$$ 

As the particles assemble on the droplet surface, the free droplet interfacial area (i.e. $A_{drop}$) decreases. With lower $A_{drop}$ due to solid particle adsorption, the rate of increase of the analyte concentration in the droplet is slower. Assuming a surface coverage of 25 \%, it would require at least 3 $\times$ 10$^{9}$ particles to cover all the droplets, estimated using the size of individual particle (i.e. 150 $nm$). In a suspension with concentration of 0.1 $mg/mL$ particles, there are 1.3 $\times$ 10$^{9}$ particles available in the capillary, which is only $\sim$ 40 \% of particles required to cover all the droplets. However, at particle concentration of 1 $mg/mL$ and above, the number of particles in the capillary is more than what is required for full droplet coverage (i.e. $\sim$ 1.2 $\times$ 10$^{10}$ particles for 1 $mg/mL$ case). This is significantly higher than the required particle number to fully cover the droplets. Therefore we can assume that all the droplets in the capillary are covered with the particles at concentrations higher than 1 $mg/mL$. This explains the slower extraction rate for suspension samples and the slight difference observed in extraction rate for concentrations higher than 1 $mg/mL$ in Fig. \ref{extraction_time}c

However, for sufficiently long time ($>$ 30 $min$) of extraction, the signal intensities of the droplets eventually become independent of the particle concentration as shown in Fig. \ref{extraction_time}d. When tested with particle concentration ranging from 3.125 to 25 $mg/mL$, the average fluorescence intensities of the droplets after extraction for more than 30 $min$ were the same (Fig. \ref{extraction_time}e). This demonstrates that within the tested range, the amount of particles do not affect the efficiency of extraction. As the blocking effect from solid particles does not change the partition coefficient, the concentration of the analyte in the droplet liquid in equilibrium with the suspension does not depend on the solid concentration. Therefore, after sufficiently long time, the analyte concentration in the droplets reached the same level, independent of the solid contents. Extraction may be influenced if the analytes and solid content in the sample interact with each other such as in the case of adsorption of chemicals onto micro/nanoplastics via hydrophobic and electrostatic interactions.\cite{tourinho2019,velzeboer2014}

\subsection{Extraction performance and sensitivity of nanoextraction from solid suspension}
\begin{figure*}[h]
	\centering
	\includegraphics[scale=1]{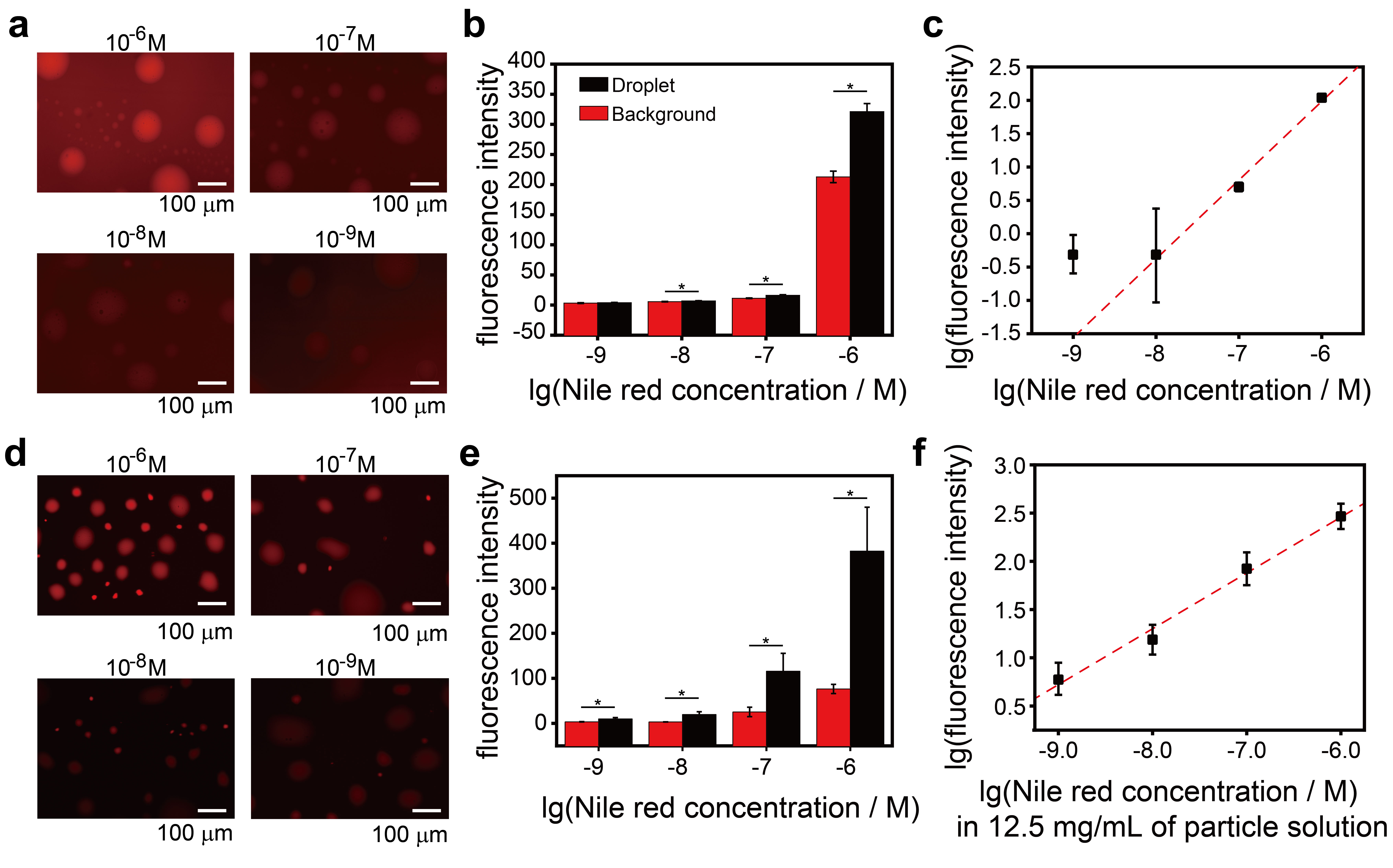}
	\caption{Limit of detection and low sample volume requirement of portable nanoextraction device. a) Fluorescent images of surface nanodroplets after extracting Nile red from water with different initial concentrations. b) Average fluorescence intensity values of surface nanodroplets and the background after extraction of Nile red from water. Here, * indicates p $<$ 0.05. c) Fluorescence intensity of droplets after subtracting the background signal. The slope in this log-log plot is $\sim$ 1.2. The outlier point indicates that the limit of detection for Nile red is 10$^{-9} M$. d) Fluorescent images of surface nanodroplets after extracting Nile red from silica solution (12.5 $mg/mL$) with different initial concentrations. e) Average fluorescence intensity values of surface nanodroplets and the background after extraction of Nile red from water. Here, * indicates p $<$ 0.05. f) Fluorescence intensity of droplets after subtracting the background signal. The slope of the log-log plot is $\sim$ 0.6} 
	\label{lod}
\end{figure*}

Fig. \ref{lod}a shows the fluorescent images of octanol surface nanodroplets after extracting Nile red dyes from particle-free aqueous solution for an initial concentrate range between 10$^{-6} M$ and 10$^{-9}  M$. Nile red has an octanol-water partition coefficient of $\lg P \sim$ 5 so the  concentration of dye in octanol is $\sim 100000\times$ that of water in equilibrium.\cite{doubell} As a result, after extraction, the droplets can clearly be differentiated from the background, owing to the higher concentration of the analyte.

Quantification of fluorescence intensities in droplets and background aqueous solution demonstrates that the limit of detection for Nile red dye from aqueous solution is 10$^{-8} M$. Fig. \ref{lod}b shows the average fluorescence intensity of the droplets and background aqueous solution measured in the vicinity of each corresponding droplet. The intensity in the droplet is higher for initial Nile red concentration range of 10$^{-6} M$ to 10$^{-8} M$ with statistical significance (p $<$ 0.05). However, at the dye concentration of 10$^{-9} M$, no difference between the droplets and background suspension was observed. Taking the logarithm of the difference between the average intensity values of droplets and of the background aqueous solution yields a calibration curve in log-log scale, correlating the fluorescence intensity and the initial dye concentration (Fig. \ref{lod}c), with a power law dependance with effective scaling exponent $\sim$ 1.2 in the range of initial concentrations between 10$^{-6} M$ to 10$^{-8} M$, which is the limit of detection.

The limit of detection depends on factors such as the partition coefficient and the quantum efficiency of the fluorescent dyes. We compared to another model compound -- Rhodamine B, which has a lower partition coefficient as compared to Nile red ($lgP \sim$ 2.3 vs $lgP \sim 5$).\cite{li2019analchem} With Rhodamine B, the limit of detection was lower (i.e. $10^{-9} M$) as compared to Nile red (Fig. S3), possibly due to its comparatively higher quantum efficiency.\cite{ghoneim2000}

In the same way, the limit of detection for a suspension sample with 12.5 $mg/mL$ of 150 $nm$ silica particles was tested using Nile red as model compound. The initial concentration of Nile red in the silica solutions varied from $10^{-6} M$ to $10^{-9} M$. Fig. \ref{lod}d shows fluorescent images of droplets after extraction of Nile red from the silica solutions. Again, the droplets can clearly be distinguished from the background suspension solution. The average fluorescence intensities of droplets are statistically higher compared to that of the background, down to initial Nile red concentration of $10^{-9} M$, which is an order of magnitude lower than in the aqueous solution. The calibration curve in log-log scale obtained from the difference in droplet and background intensities reveals a power law, but with a lower effective scaling exponent of $\sim$ 0.6. The lower limit of detection and lower effective scaling exponent in the calibration curve may be due to the interaction of Nile red molecules at the particle-water interface.

\begin{figure}[h]
	\centering
	\includegraphics[scale=1]{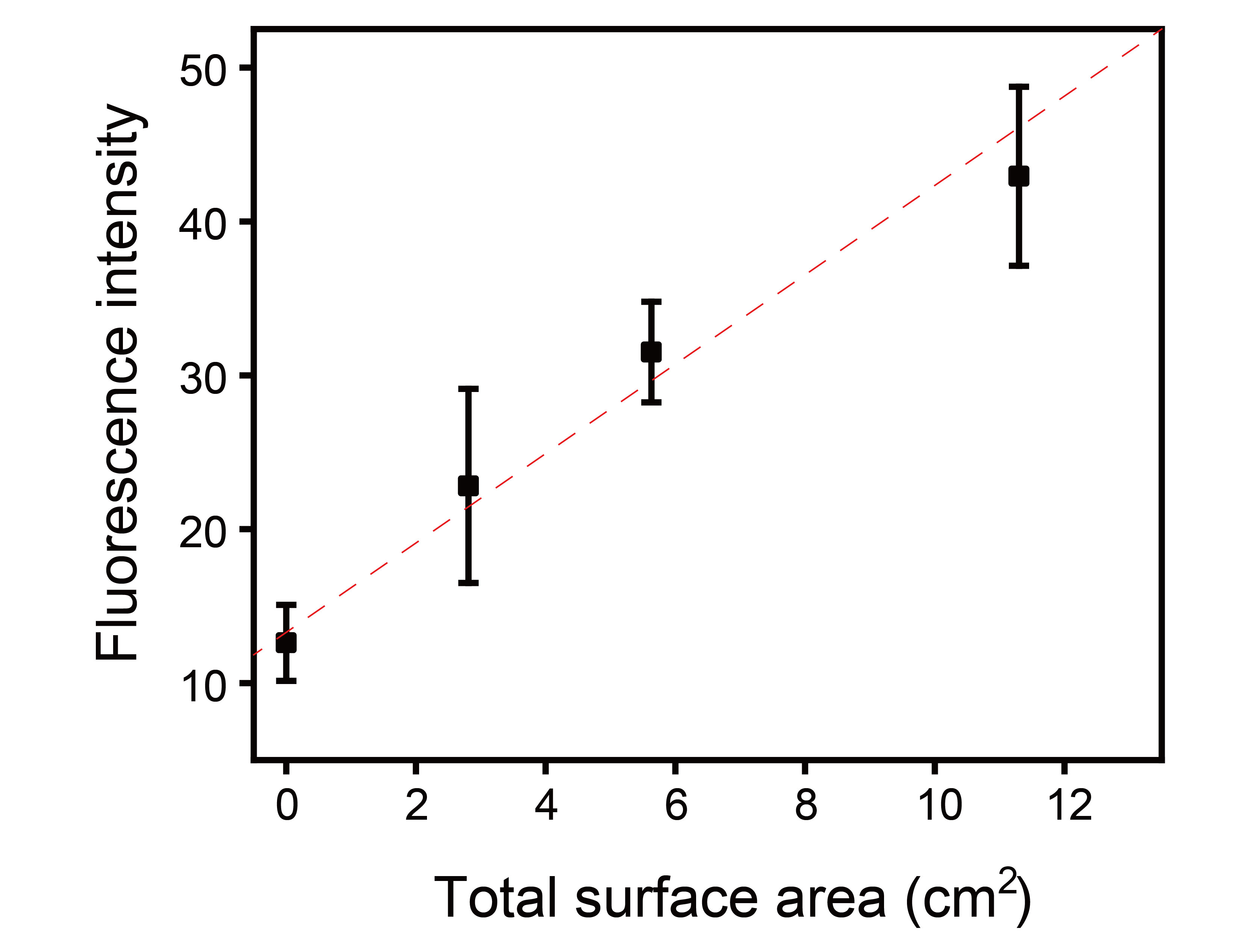}
	\caption{Plot showing the relationship between the intensity of the droplet with the total surface area offered by the particles. The total surface area of the particles was calculated by multiplying the surface area of a 150 $nm$ particle with the total number of the particles available in the capillary tube.}
	\label{surface area}
\end{figure}

The advantageous high fluorescence intensity in the suspension samples compared to aqueous samples may be attributed to the adsorption and accumulation of the dye molecules at particle-liquid interfaces. Fig. \ref{surface area} shows a plot of the fluorescence intensity of droplets and total surface area of particles corresponding to the concentration range between 3.125 $mg/mL$ to 12.5 $mg/mL$. The total surface area of particles was calculated by multiplying the surface area of a single 150 $nm$ particle (density $\sim$ 2 $g/cm^{3}$) with the total number of particles available in the capillary tube (volume $\sim$ 48 $\mu L$), which is determined by the particle concentration in suspension. At higher particle concentration, there is more area available for the analyte molecules to adsorb such that with more particles adsorbed at droplet-water interface, more analyte is available as well to increase the fluorescence signal.

\subsection{Influence of sample volume on extraction performance}

\begin{figure*}[h]
	\centering
	\includegraphics[scale=1]{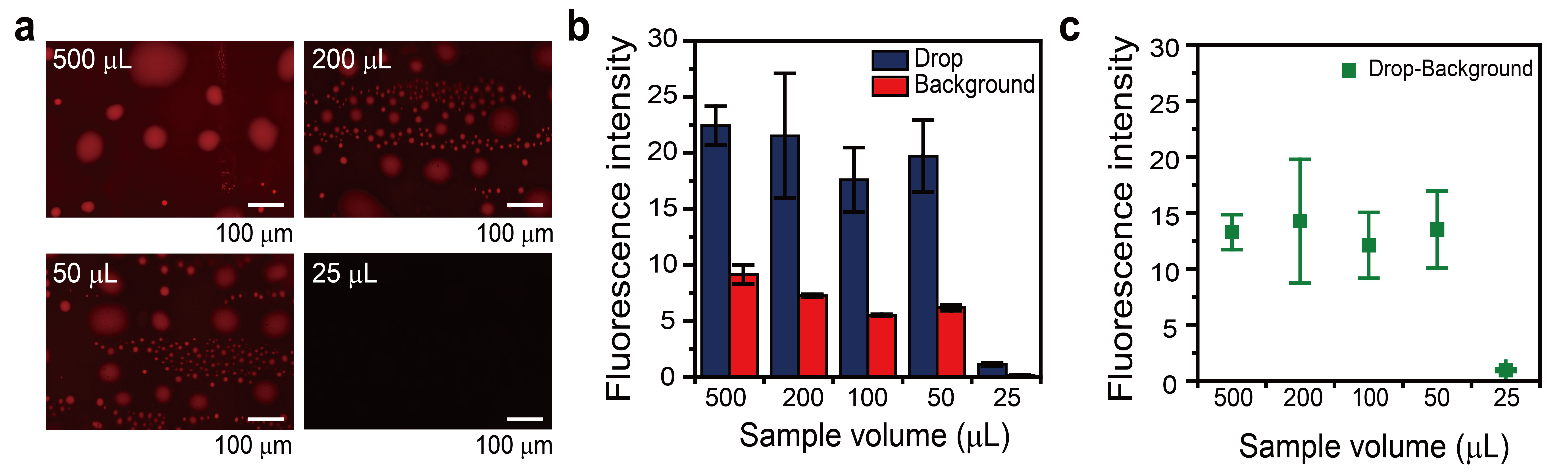}
	\caption{Processing of sample at low volume. a) Fluorescence images of surface nanodroplets after extracting a 1 $\mu$M aqueous solution of Nile red at different sample volumes.  For better visualization purposes, the brightness has been adjusted. b) Fluorescence intensity values of nanodroplets and background after extraction of 10$^{-6} M$ Rhodamine B solution. c) Fluorescence intensity of droplets after subtracting the background signal. The intensity values for various sample volumes are similar from 500 $\mu$L to 50 $\mu$L, but the signal is negligible at sample volume of 25 $\mu$L.}
	\label{volume}
\end{figure*}

Apart from simplicity and high efficiency, another crucial advantageous feature of nanoextraction is the small sample volume (on the order of $\mu L$), thanks to the confined geometry of the glass capillary. As shown in Fig. \ref{volume}a-b, detection of Nile red (concentration $=$ 10$^{-6} M$) can be performed with sample volumes as low as 50 $\mu L$. Fluorescence intensity of both the droplets and background remain unchanged down to 50 $\mu L$ (Fig. \ref{volume}b). As a result, the difference between the signal from droplet and that from the background is also similar for the same range of sample volumes (Fig. \ref{volume}c), demonstrating reliable extraction. 

The reason reliable extraction is observed with volume as low as 50 $\mu L$ is attributed to the volume of the capillary, which is $\sim$ 48 $\mu L$ (i.d. = 1.1 $mm$ and length = 50 $mm$). When the sample is introduced to the capillary, it can completely replace solution B and the droplets can extract the target analytes. On the other hand, for the case of 25 $\mu L$ of volume, the sample does not thoroughly replace the solution contained in the droplet B such that the fluorescent intensity is very low when imaged with the same exposure time. 

In addition, the fact that surface nanodroplets are on the scale of femtoliters plays a significant role in lowering the required sample volume.\cite{li2019small} Liquid-liquid extraction relies on the partition coefficient ($\lg$ P) of the analytes which is determined by the concentration of the analytes in the sample and the extractant oil at equilibrium. Thus, even if the sample volume is low, the volumes of surface nanodroplets can still be much lower than that of the sample, enabling reliable extraction.

Extraction of analytes from a low sample volume is of high interest in many diagnostic applications involving body fluids such as saliva\cite{jouyban2019, timofeeva2016} or blood.\cite{odoardi2015} However, in conventional extraction methods such as DLLME or single-drop microextraction, the sample volume is usually increased by diluting the sample with water prior to extraction. In doing so, the concentration of analyte in the sample decreases, reducing the sensitivity. Moreover, many of the commonly used microextraction processes that employ centrifugation are challenged by the difficulty of separating the centrifuged oil pellet from the sample if the sample volume is low. 

\subsection{Extraction of target analyte from oil sand wastewater}
The portability of the entire device is potentially useful for analysis in the field such as in environmental monitoring. As a proof-of-concept demonstration, we show extraction of target analyte from oil sand wastewater comprising of solid particles, bitumen (heavy oil) and other hazardous hydrocarbons such as naphthenic acids or adamantane.\cite{hewitt2020} Detection of such analytes from oil sand wastewater is important for the environment as seepage of these chemicals into the ground or surface waters can cause adverse effects to the aquatic life.\cite{hewitt2020}

\begin{figure}[h]
	\centering
	\includegraphics[scale = 1]{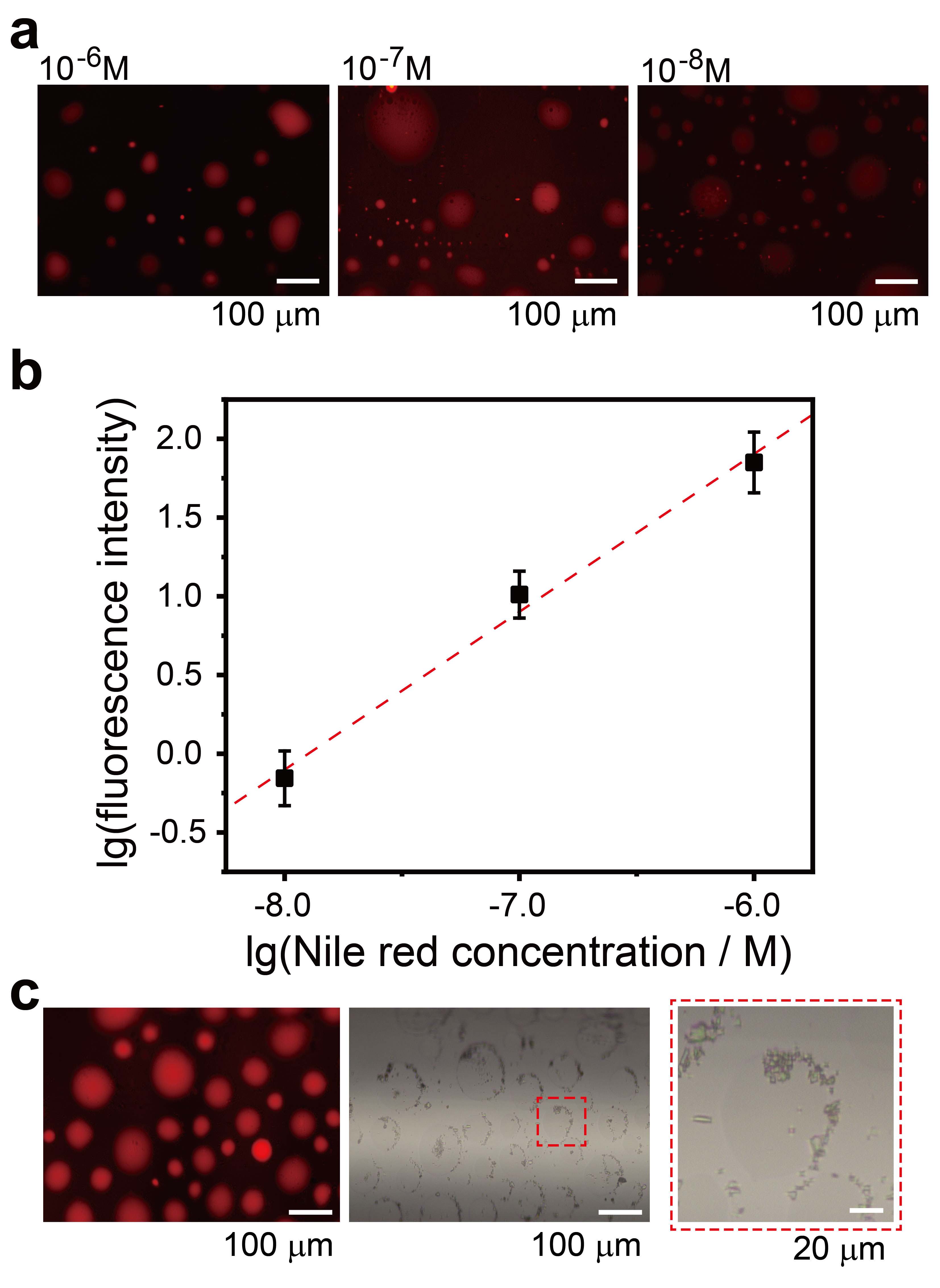}
	\caption{Extraction of analyte from complex oil sand wastewater sample. a) Fluorescence images of oil sand wastewater with Nile red at $10^{-6} M$ to $10^{-8} M$. b) Limit of detection for Nile red in oil sand wastewater sample. When plotted in log-log scale, a linear trend is observed, i.e. the slope is $\sim$1. c) Extraction of 10$^{-6} M$ Nile red dye from oil sand wastewater with high solid content ($\sim$ 30 wt\%). The dye is readily extracted into the droplets. However, presence of particle aggregates are observed at the droplet-water interface.}
	\label{tailing}
\end{figure}

In Fig. \ref{tailing}a we show extraction of Nile red from a oil sand wastewater sample comprising of 0.1 wt\% bitumen, 7 wt\% solids (fines and sands) and 92.9 wt\% water. Prior to extraction, Nile red -- mimicking hydrophobic compounds in the wastewater -- was added to the oil sand wastewater sample at concentrations ranging from $10^{-6} M$ to $10^{-9} M$. Similar to extraction from silica suspension, we determined the limit of detection for the oil sand wastewater sample based on fluorescence intensity in the droplet and in the background. The limit of detection was $10^{-8} M$, an order of magnitude lower than that of silica solution as shown in Fig. \ref{tailing}b. The lower detection sensitivity from oil sand wastewater may be attributed to the sorption of Nile red molecules to the solid particles that are potentially fouled by bitumen. As bitumen is known to be an aggressive foulant,\cite{chen2017}  it is possible for solid particles to have bitumen coating on the surface. Then Nile red can be adsorbed on the bitumen-coated solids thereby lowering the liquid-liquid extraction efficiency by the surface nanodroplets.

Nonetheless, extraction was successful even with an oil sand wastewater sample with a higher solid content of $\sim$ 30 wt\% as shown in Fig. \ref{tailing}c. The sample was spiked with 10$^{-6} M$ of Nile red dye and it was injected to the capillary device after forming surface nanodroplets with octanol. Nile red dye was readily extracted into the droplets from which strong fluorescence signal was detectable. Although some particles aggregated at the droplet-water interface due to their excessive amount in the sample, their influence would be minimal on the detection of extracted analytes in the droplets by in-situ methods.

\section{Conclusion}
In summary, we developed a gree and sustainable nanoextraction approach for concentrating compounds from highly concentrated solid suspensions using surface nanodroplets. Extraction of model analytes was feasible for suspension samples with solid content up to 30 wt\%. Without prior removal of solids, nanoextraction is fast, requiring only sub-milliliter sample volume. As a proof-of-concept, a target compound was extracted from an oil-sand slurry. A similar limit of detection was observed for both aqueous samples and suspension samples ($10^{-8} M \sim 10^{-9} M$), demonstrating that particles do not reduce the extraction efficiency of the target analytes. Instead, the particles initially slowed down the extraction rate due to their adsorption to the droplets.

In future work, the extracted analytes can be analyzed using other in-situ methods such as UV-Vis microphotospectrometry, surface-enhanced Raman spectroscopy, or attenuated total reflection (ATR) infrared spectroscopy. We expect that the technology shown in this work provides opportunities in rapidly and easily separating target analytes from complex samples in low volumes such as saliva or blood, which will be useful for diagnostic applications, or for online analysis of slurries in the environment.

\section*{Conflicts of interest}
There are no conflicts to declare.

\section*{Acknowledgements}
This project was supported by the ERC Proof-of-Concept grant (Project Number 862032). The project was also partly supported by the Natural Science and Engineering Research Council of Canada (NSERC) and Future Energy Systems (Canada First Research Excellence Fund). XZ acknowledges support from the Canada Research Chairs Program.



\balance


\bibliography{ref} 
\bibliographystyle{rsc} 

\end{document}


\begin{doublespace}

\maketitle

\setcounter{figure}{0}
\makeatletter 
\renewcommand{\thefigure}{S\arabic{figure}}
\begin{figure}
    \centering
    \includegraphics[scale=1]{./Figures/steps(SI).jpg}
    \captionsetup{font = {small}}
    \caption{Steps for operating portable nanoextraction device for nanodroplet forming and extraction.}
    \label{steps}
\end{figure}

\begin{figure}
    \centering
    \includegraphics[scale=1]{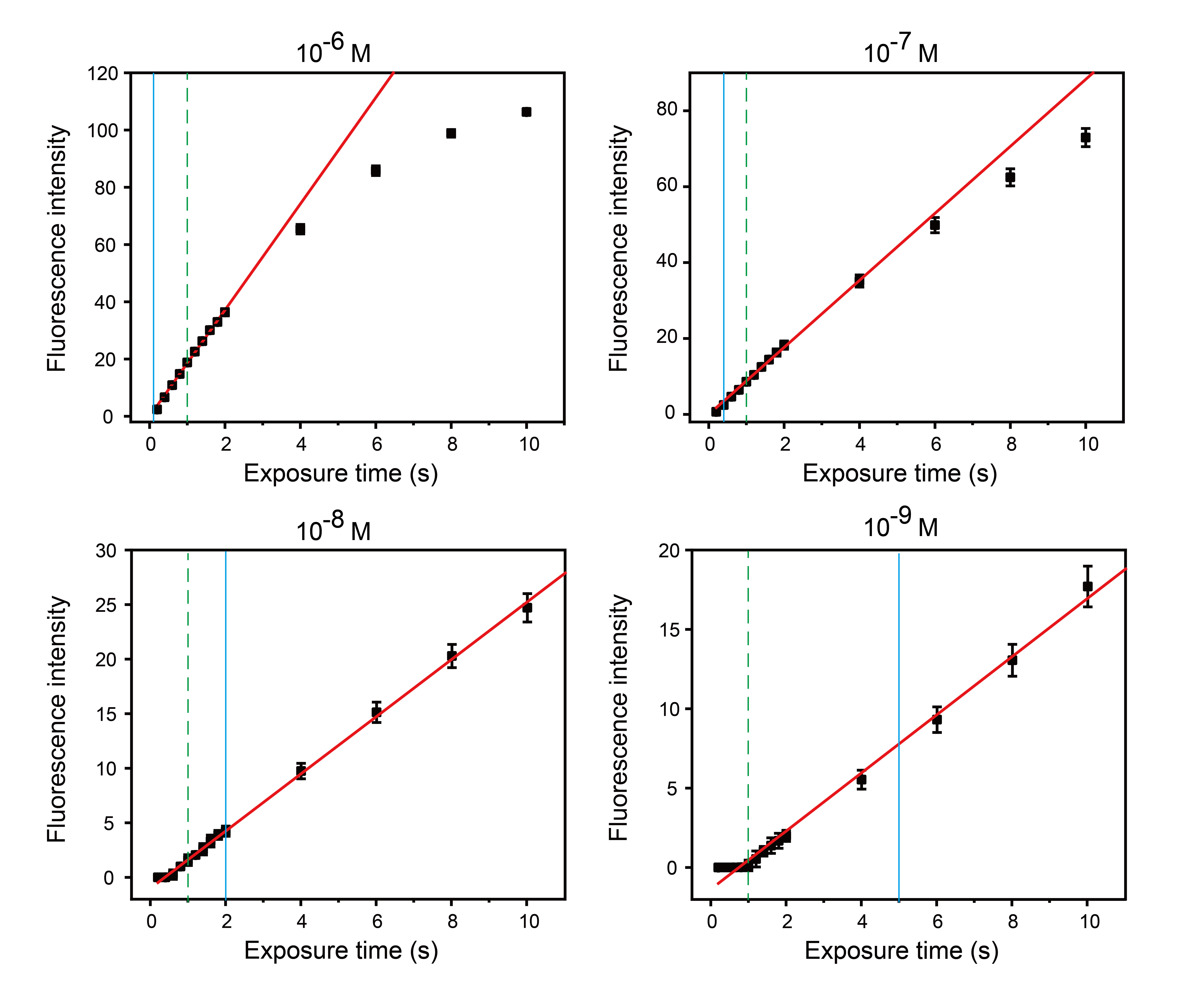}
    \captionsetup{font = {small}}
    \caption{Influence of exposure time on background Nile red fluorescence intensity for 10$^{-6} M$ to 10$^{-9} M$. Red dashed line (\textcolor{redcolor}{\textbf{- - -}}) indicates the best fit line for linear region of fluorescence. The green dotted line (\textcolor{greencolor}{\textbf{$\cdot \cdot \cdot$}}) represents the exposure time used for each experiment. Blue solid line (\textcolor{bluecolor}{\textbf{---}}) shows the exposure time to which the intensity data were normalized to (t = 1 $s$)}
    \label{background}
\end{figure}

\begin{figure}
    \centering
    \includegraphics[scale=1]{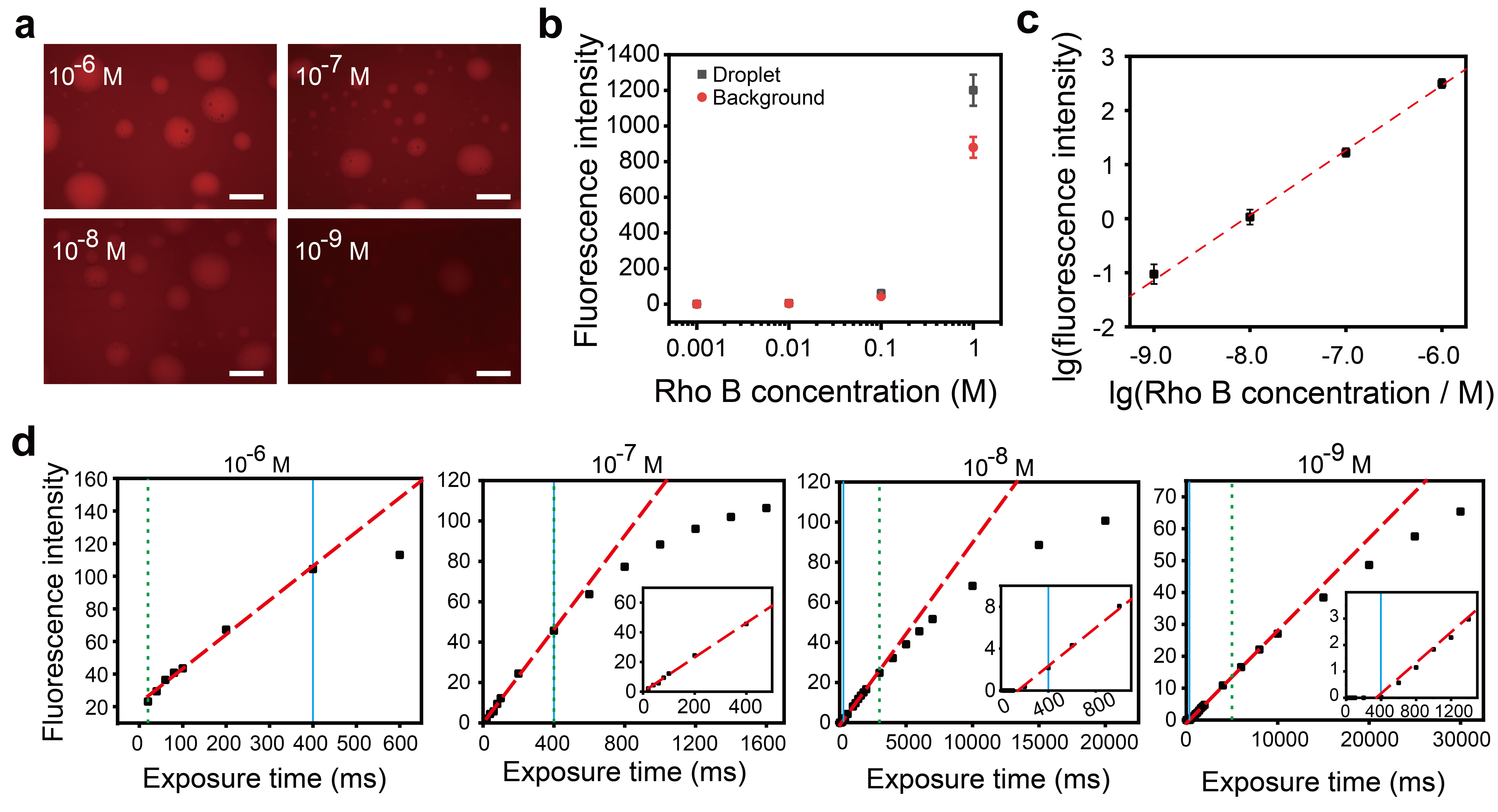}
    \captionsetup{font = {small}}
    \caption{Limit of detection for Rhodamine B dye. a) Images of surface nanodroplets after extracting Nile red from different initial concentrations. For better visualization purposes, the brightness has been adjusted. b) Average fluorescence intensity values of surface nanodroplets as well as the background after nanoextraction for various initial concentration of Nile red. c) Fluorescence intensity of droplets after subtracting the background signal. The trend has a slope $\sim$1.2 when plotted in log-log scale. d) Background fluorescence intensities for Rhodamine B solution at various exposure times ranging 0 to 30 $s$. Blue solid line (\textbf{\textcolor{bluecolor}{---}}) indicates the exposure time to which the intensity data were normalized to (t = 400 $ms$)} 
    \label{LOD_RhoB}
\end{figure}

\end{doublespace}